\documentclass[arXiv]{actapoly}
\usepackage{graphicx}
\begin{document}

\title[Example of an Article with a Long Title]
{Proton stopping power of different density profile plasmas}

\correspondingauthor[D. Casas]{David Casas}{my}{David.Casas2@alu.uclm.es}
\author[M.D. Barriga-Carrasco]{Manuel D. Barriga-Carrasco}{my}
\author[A.A. Andreev]{Alexander A. Andreev}{their}
\author[M. Schnürer]{Matthias Schn\"urer}{their}
\author[R. Morales] {Roberto Morales}{my}

\institution{my}{E.T.S.I. Industriales, Universidad de Castilla-La Mancha, Av. Camilo Jos\'e Cela s/n 13071, Ciudad Real, Spain}
\institution{their}{Max Born Insitute, Max Born Str. 2a D-12489, Berlin, Germany}

\begin{abstract}
In this work, the stopping power of a partially ionized plasma is analyzed by means of free electron stopping and bound electron stopping. For the first one, the RPA dielectric function is used, and for the latter one, an interpolation of high and low projectile velocity formulas is used. The dynamical energy loss of an ion beam inside a plasma is estimated by using an iterative scheme of calculation. The Abel inversion is also applied when we have a plasma with radial symmetry. Finally, we compare our methods with two kind of plasmas. In the first one, we estimate the energy loss in a plasma created by a laser prepulse, whose density is approximated by a piecewise function. For the latter one, a radial electron density is supposed and the stopping is obtained as function of radius from the calculated lateral points. In both cases, the dependence with the density profile is observed.
\end{abstract}


\maketitle

\section{Introduction}

Nuclear fusion has a promising future as clean, endless, and sustainable energy source for humankind. A large amount of energy is achieved from a dense and highly energetic deuterium-tritium plasma. However, there are great challenges in order to obtain this so dense and overheated plasma. One of the chosen methods is by means of energetic beams as high power lasers or fast particles. In this last case, it is important to study energy loss of an ion beam that passes through a plasma target to understand the interactions of swift particles with nuclear fusion fuel pellet.

\begin{figure}
\centering
\includegraphics[width=\linewidth]{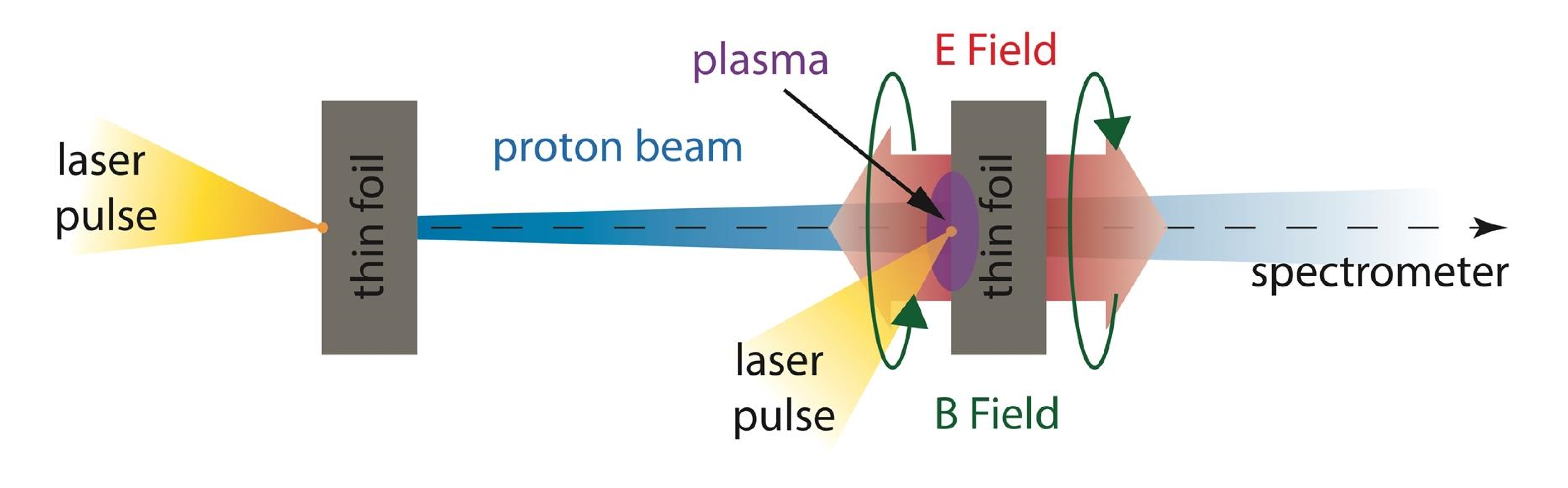} 
\caption{Sketch of proton beam interaction with plasma target.}
\label{F01}
\end{figure}

The proton is the lightest ion that can be accelerated, and it achieves a great velocity due to high ratio charge-mass. Furthermore, the laser-accelerated proton beams technique has achieved a great development in the last years \cite{Macchi2013,Daido2012}. The low longitudinal emittance of the beam together with a continuous distribution of proton kinetic energies of a few MeV allow to trace the temporal evolution of strong electric and magnetic fields in plasma foils \cite{Abicht2013}. A diagram of this process is shown in Figure \ref{F01}. The temporal evolution of energy loss can be evaluated using the proton streak deflectometry, where the proton energy, which encodes the time, is resolved using a magnetic spectrometer.

Electronic stopping is the main processes that contributes to deposit energy on plasma target for ion or proton beams. For partially ionized plasmas, this stopping is divided into two contributions: free electrons and bound electrons. Both are calculated using different methods: For the first one, the random phase approximation (RPA) dielectric function is used, and for the latter one, an interpolation formula between limits of high and low projectile velocities together with Hartree-Fock calculations for atomic quantities is utilized. \cite{CASASPRE2013,Barriga-Carrasco2013LPB}.

Stopping power calculation methods are described in section \ref{Theory}. Afterwards, stopping power of different density profile plasmas are estimated for a proton beam in section \ref{Results} and finally a summary of this work is explained in section \ref{Conclusions}. We will use atomic units (a.u.) $e=\hbar=m_{e}=1$ to simplify expressions.

\section{Theoretical Methods}\label{Theory}

\subsection{Free Electron Stopping}

Stopping power of free plasma electrons can be calculated using a RPA dielectric function (DF). In this DF, the effect of a swift charged particle that passes through an electron gas is considered as a perturbation that losses energy proportionally to the square of its charge. Then slowing-down was simplified to a treatment of the properties of the medium only, and a linear description of these properties may then be applied.

The RPA dielectric function (DF) is developed in terms of the wave number $k$ and of the frequency $\omega$ provided by a consistent quantum mechanical analysis. The RPA analysis yields the expression \cite{Lindhard54}
\begin{equation}\label{EpsRPA}
\varepsilon_{{\rm RPA}}(k, \omega)=1+\frac{1}{\pi^{2}k^{2}}\int\!\! d^{3}k'
\frac{f(\overrightarrow{k}+\overrightarrow{k}')-f(\overrightarrow{k}')}{\omega+i\nu-(E_{\overrightarrow{k}+\overrightarrow{k}'}-E_{\overrightarrow{k}'})},
\end{equation}
where $E_{\overrightarrow{k}}=k^{2}/2$. The temperature dependence is included through the Fermi-Dirac function
\begin{equation*}
f(\overrightarrow{k})=\frac{1}{1+\exp[\beta(E_{k}-\mu)]},
\end{equation*}
being $\beta=1/k_{B}T$ and $\mu$ the chemical potential of the plasma with electron density $n_{e}$ and temperature $T$. In this part of the analysis, we assume the absence of collisions so that the collision frequency tends to zero, $\nu \rightarrow 0$.

The analytic RPA dielectric function for plasmas at any degeneracy can be obtained directly from Eq. (\ref{EpsRPA}) \cite{Gouedard1978,Arista1984}:
\begin{equation}\label{EpsRPAnew}
\varepsilon_{{\rm RPA}}(k, \omega)=1+\frac{1}{4z^{3}\pi k_{F}}[g(u+z)-g(u-z)],
\end{equation}
where $g(x)$ corresponds to
\begin{equation*}
g(x)=\int^{\infty}_{0}\frac{ydy}{\exp(Dy^{2}-\beta\mu)+1}\ln \left( \frac{x+y}{x-y} \right) ;
\end{equation*}
$u=\omega / kv_{F}$ and $z=k/2k_{F}$ are the common dimensionless variables \cite{Lindhard54}. $D=E_{F}\beta$ is the degeneracy parameter and $v_{F}=k_{F}=\sqrt{2E_{F}}$ is Fermi velocity in a.u.

Finally, electronic stopping of free plasma electrons will be calculated in the dielectric formalism as
\begin{equation}\label{SpFree}
Sp_{f}(v)=\frac{2Z^{2}}{\pi v^{2}}\int^{\infty}_{0}\!\!\frac{dk}{k}\int^{kv}_{0}\!\!d\omega\,\omega\,{\rm Im}
\left[ \frac{-1}{\varepsilon_{{\rm RPA}}(k, \omega)} \right],
\end{equation}
where $Z_{p}$ is the charge, $v$ is the velocity of the projectile, and the equation is in atomic units.

\begin{figure}
\centering
\includegraphics[width=\linewidth]{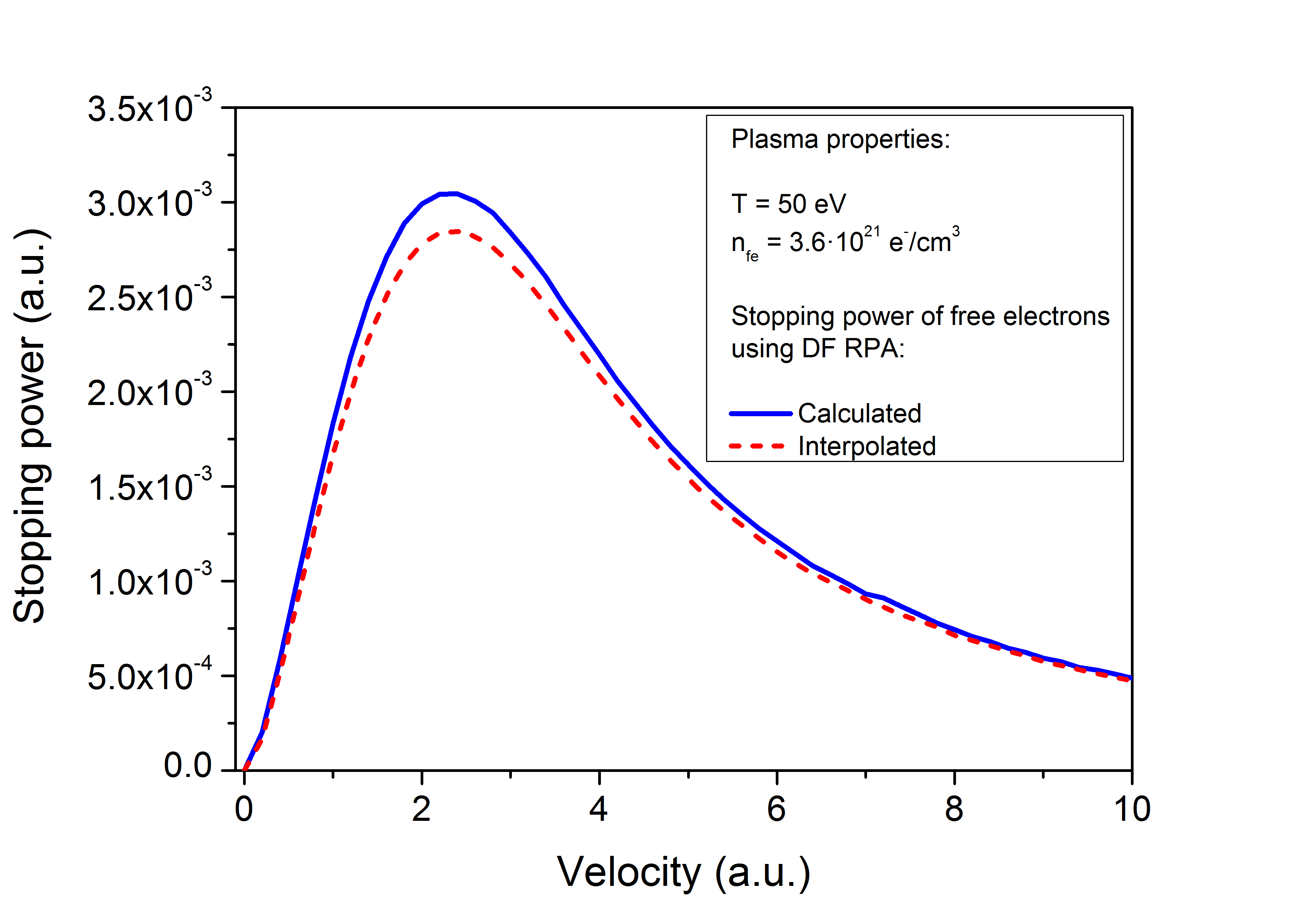} 
\caption{Stopping power as a function of proton beam velocity. Blue solid line: Direct calculation of RPA. Red dashed line: Interpolation of RPA}
\label{F02}
\end{figure}

The calculation of stopping power using (\ref{SpFree}) could be difficult and computationally slow in some cases. A fast accurate method is to make an interpolation from a database \cite{PELO2013} where the variables to interpolate are temperature and density for every couple of coordinates $(v_{i},Sp_{i})$ located on the interpolation grid. Database has a set of RPA result files for different conditions of temperature and density. For sake of simplicity, bilinear method is used for a 2D interpolation, which is an extension of linear interpolation for interpolating functions of two variables on a regular 2D grid. In Figure \ref{F02}, we can see the difference between the direct calculation of RPA and its interpolation, for a free electron density of $3.6\times10^{21}\,\,{\rm e^{-}/cm^{3}}$. Both graphs are similar with a slight difference on the maximum values of the stopping, where the interpolated RPA is lower that the calculated one.

\subsection{Bound Electron Stopping}
The stopping power of a cold gas or a plasma for an ion with charge $Z$ has been calculated many times by the well-known Bethe formula \cite{Bethe}
\begin{equation}\label{BetheF}
Sp=\left[\frac{Ze\omega_{p}}{v_{p}}\right]^2\ln\left[\frac{2m_{e}v^{2}_{p}}{I}\right],
\end{equation}
where $\omega^{2}_{p}=4\pi n_{e}e^{2}/m_{e}$ is the square of the plasma frequency and $n_{e}$ denotes the bound electron density. $I$ is the mean excitation energy, which averages all the exchanged energy in excitation and/or ionization processes between a fast charged particle and the target bound electrons. We can simplify (\ref{BetheF}) using atomic units
\begin{equation}\label{BetheAU}
Sp=\left[\frac{Z\omega_{p}}{v_{p}}\right]^2\ln\left[\frac{2v^{2}_{p}}{I}\right].
\end{equation}
$I$, is estimated by several methods. A short expression is deduced in ref. \cite{Garbet1987}
\begin{equation}\label{MEE}
I=\sqrt{\frac{2K}{\langle r^{2} \rangle}},
\end{equation}
where $K$ is electron kinetic energy and $\langle r^{2} \rangle$ is the average of the square of the radius. These quantities can be estimated for the whole atom/ion or shell by shell using atomic calculations.

However, Bethe equation has a disadvantage, when the logarithm argument in Eq. (\ref{BetheAU}) is less than one, this results in a negative stopping, which has not physical meaning. To avoid this difficulty, an interpolation formula obtained in ref. \cite{Barriga2005LPB} is used in this work. This expression interpolates the stopping between the limits of high velocity and low velocity projectiles, which are separated by an intermediate velocity
\begin{equation}\label{BoundInter}
L_{b}(v)=\left\{ \begin{array}{lr}
L_{H}(v)=\ln\frac{2v^{2}}{I} - \frac{2K}{v^{2}} & {\rm for }\; v>v_{{\rm int}}
\\L_{B}(v)=\frac{\alpha v^{3}}{1+Gv^{2}}& {\rm for }\; v\leq v_{{\rm int}}
\end{array} \right.
\end{equation}
\begin{equation}\label{Vint}
v_{{\rm int}}=\sqrt{3K+1.5I},
\end{equation}
where $G$ is given by $L_{H}({v_{\rm int}}) = L_{B}({v_{\rm int}})$, and $\alpha$ is the friction coefficient for low velocities. $L_{b}(v)$ substitutes the logarithm in Eq. (\ref{BetheAU}). Using equations (\ref{BetheAU}) and (\ref{BoundInter}), the stopping power of bound electrons for a proton beam ($Z=1$) is
\begin{equation}\label{SpBound}
Sp_{b}=\frac{4\pi n_{at}}{v^{2}}L_{b}(v)
\end{equation}

\subsection{Energy Loss in a Thick Plasma Target}
\begin{figure}
\centering
\includegraphics[width=\linewidth]{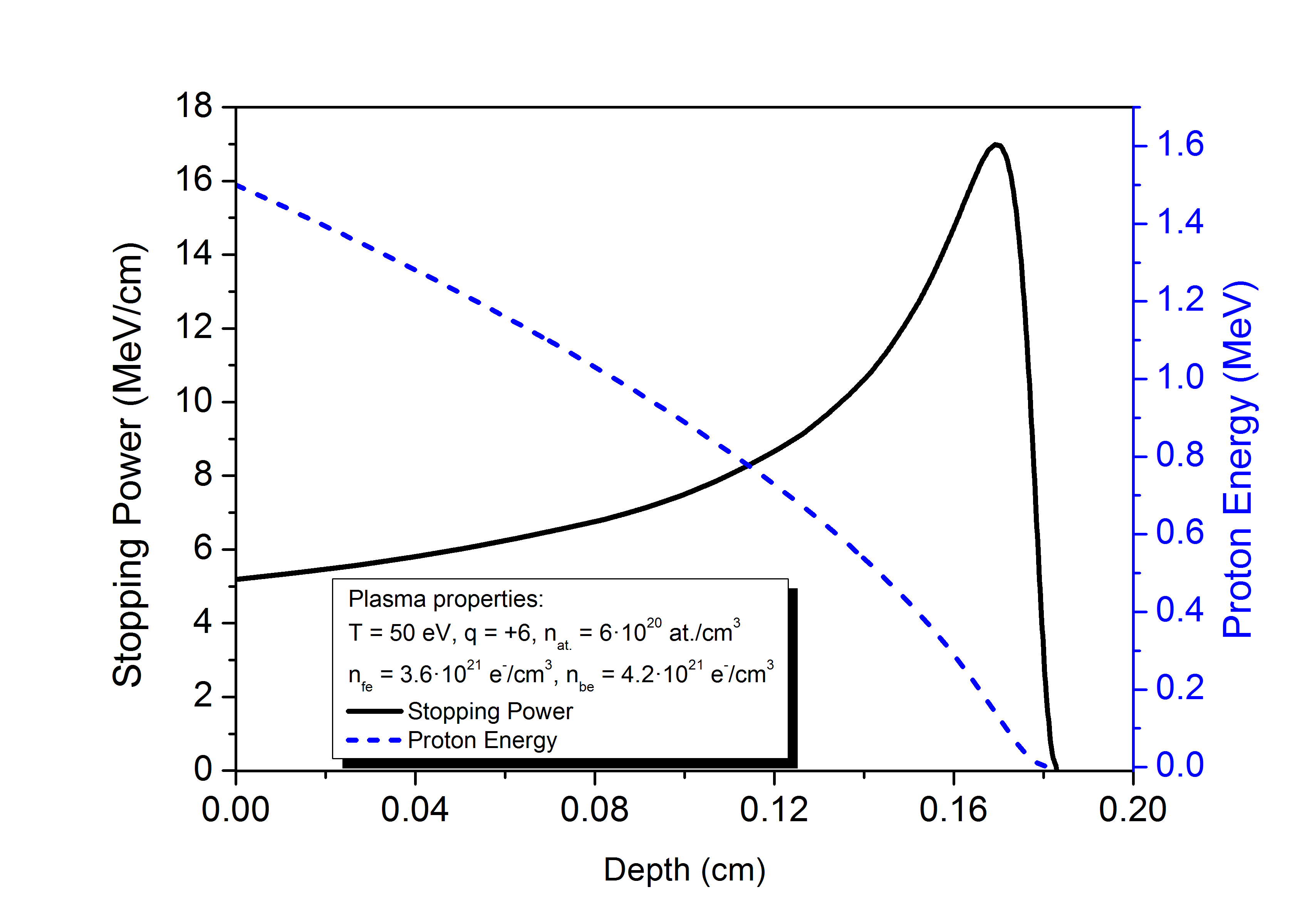} 
\caption{Stopping power (thick line) and proton energy (dashed line) as a function of depth for a 1.5 MeV proton beam. The Bragg peak is placed at 0.169 cm.}
\label{F03}
\end{figure}

The energy loss of a proton beam in a material, like plasma, is a dynamical process. When it impacts with an initial energy, $E_{p_{0}}$, it starts losing energy with a rate that is given by the stopping power function. Using an iterative scheme, this energy loss could be calculated. The method is to divide the plasma length in segments and to evaluate the energy loss in the ith step by means of
\begin{equation*}
E_{L_{i}}=\frac{Sp_{i}}{\Delta x},
\end{equation*}
where $Sp_{i}$ is the stopping in the ith segment and $\Delta x$ its length.
Applying this iterative calculation to a plasma stopping profile, it is possible to calculate the energy loss profile and the Bragg peak for a proton beam that is totally stopped inside the target. In Figure \ref{F03}, both graphs are been calculated for a partially ionized aluminum plasma.

\subsection{Abel Inversion}
\begin{figure}
\centering
\includegraphics[width=\linewidth]{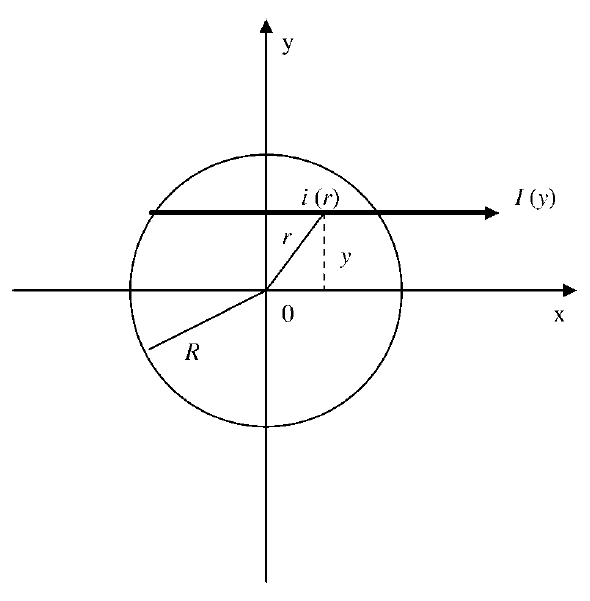} 
\caption{The radial distribution $i(r)$ cannot be measured directly, but only through the integral $I(y)$ in the $x$-direction.}
\label{F04}
\end{figure}

The Abel inversion method is a mathematical technique that has been used to analyze proton imaging data from inertial confinement fusion experiments \cite{DeCiantis2006,Seguin2006}. With this technique a set of radial points is obtained from a corresponding set of lateral data points. The relationship between the lateral intensity measured $I(y)$ and the radial intensity desired $i(r)$ is shown schematically in Fig. \ref{F04} and is given by
\begin{equation}\label{Abel01}
I(y)=\int^{x_{0}}_{-x_{0}} i(r)dx
\end{equation}

In this expression the integral is taken along a strip at constant $y$, $x^{2}+y^{2}=r^{2}$, $x_{0}^{2}=R^{2}-y^{2}$ is the $x$ coordinate of the plasma edge at $y$ value, and $R$ is the radius beyond which $i(r)$ is negligible. Hence, assuming a radial symmetry, Eq. (\ref{Abel01}) can also be written
\begin{equation}\label{Abel02}
I(y)=2\int^{R}_{y}i(r)\frac{r}{\sqrt{r^{2}-y^{2}}}dr
\end{equation}

Equation (\ref{Abel02}) is one form of Abel$'$s integral equation. The reconstruction of the unknown function $i(r)$ from the measured data $I(y)$ can be done analytically by means of the inverse of Abel$'$s integral equation
\begin{equation}\label{Abel03}
i(r)=-\frac{1}{\pi}\int^{R}_{r}\frac{dI(y)}{dy}\frac{dy}{\sqrt{y^{2}-r^{2}}}
\end{equation}

The experimental measurement of $I(y)$ provides a discrete set of data points. Thus, both the differentiation and the integration in Eq. (\ref{Abel02}) cannot be performed directly. For this reason, the Nestor-Olsen method \cite{NestorOlsen60} is used in section \ref{Results} to apply the Abel inversion to a discrete set of stopping power points.

\section{Results}\label{Results}

\begin{figure}
\centering
\includegraphics[width=\linewidth]{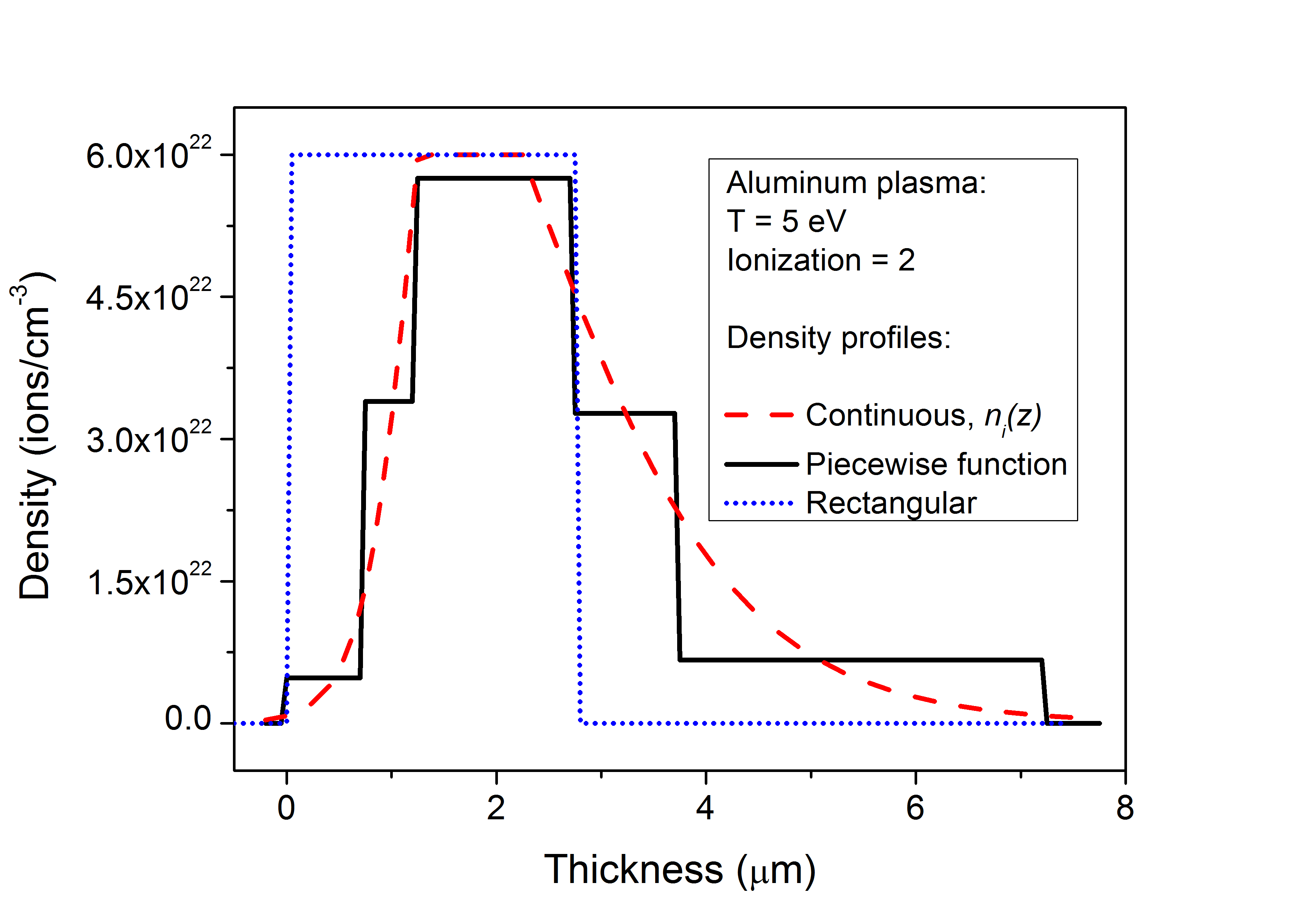} 
\\[3mm]
\includegraphics[width=\linewidth]{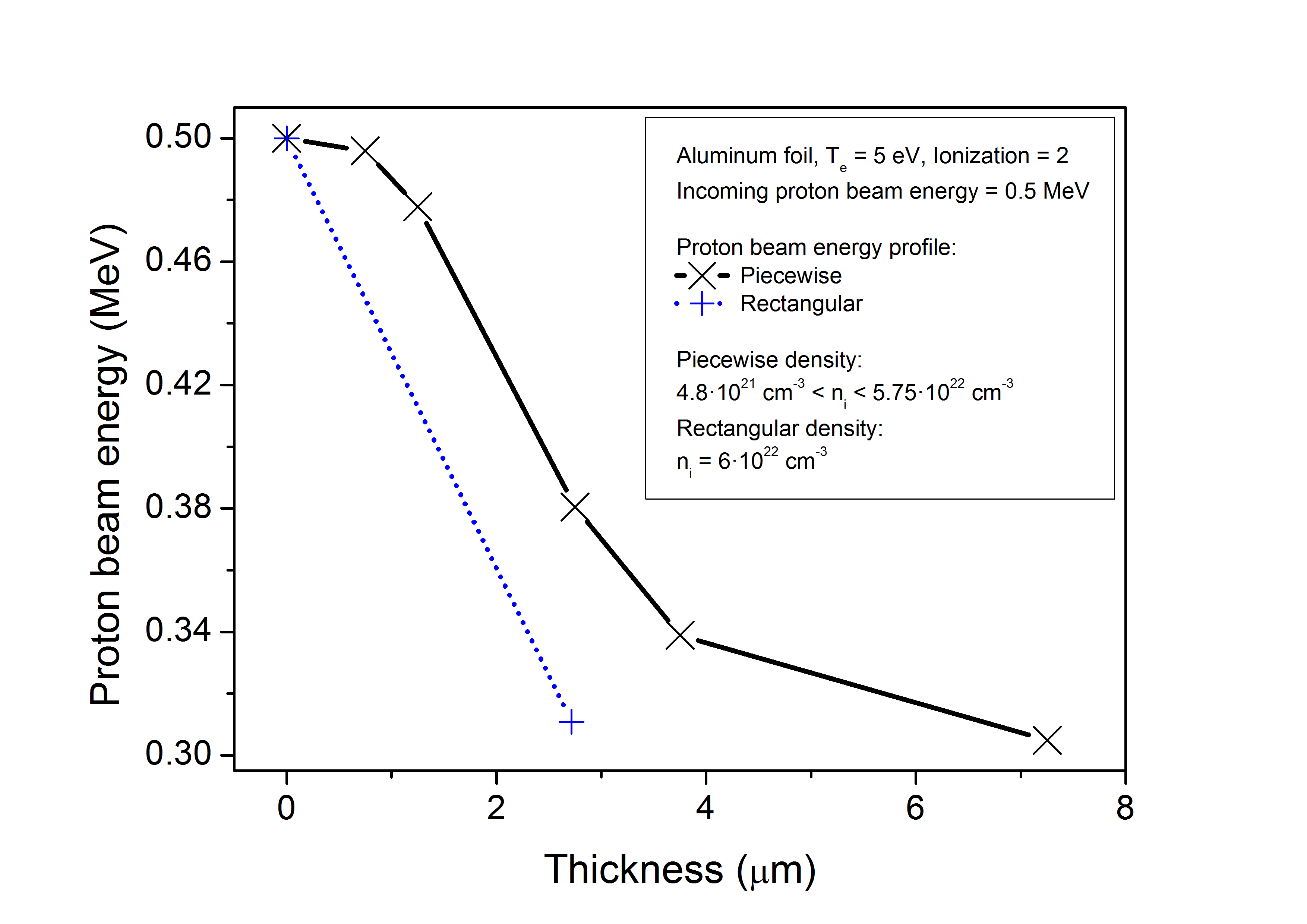} 
\caption{The target density profiles (top) and its corresponding energy loss functions (bottom).}
\label{F05}
\end{figure}

Using stopping power expressions, (\ref{SpFree}) and (\ref{SpBound}), is possible to estimate the energy loss of a proton beam for different density target distributions: rectangular shape with a constant density and the piecewise approximation of a trapezium shape with a density profile given by \cite{Andreev2009}
\begin{equation}\label{Andreev}
n_{i}(z)=\frac{2n_{i{\rm max}}}{1+\exp\left[ \frac{2x\theta(x)}{l_{r}}-\frac{2x\theta(-x)}{l_{fr}} \right]]}.
\end{equation}
The Eq. (\ref{Andreev}) is the density distribution obtained when a laser prepulse hit a thin target with a thickness of 1 micron or less. Here $x=z-0.5l_{r}$ and $\theta(x)$ is the Heaviside step function. The parameters $l_{r}$ and $l_{fr}$ were obtained for different initial target thicknesses, $l_{f}$, from hydrocode calculations \cite{Andreev2009}. $l_{r}$, $l_{fr}$, and $l_{f}$ are expressed in microns. For a solid aluminum target, $n_{i{\rm max}}=6\times10^{22}cm^{-3}$. The different density profiles and the corresponding proton beam energy are showed in Figure \ref{F05}.

\begin{figure}
\centering
\includegraphics[width=\linewidth]{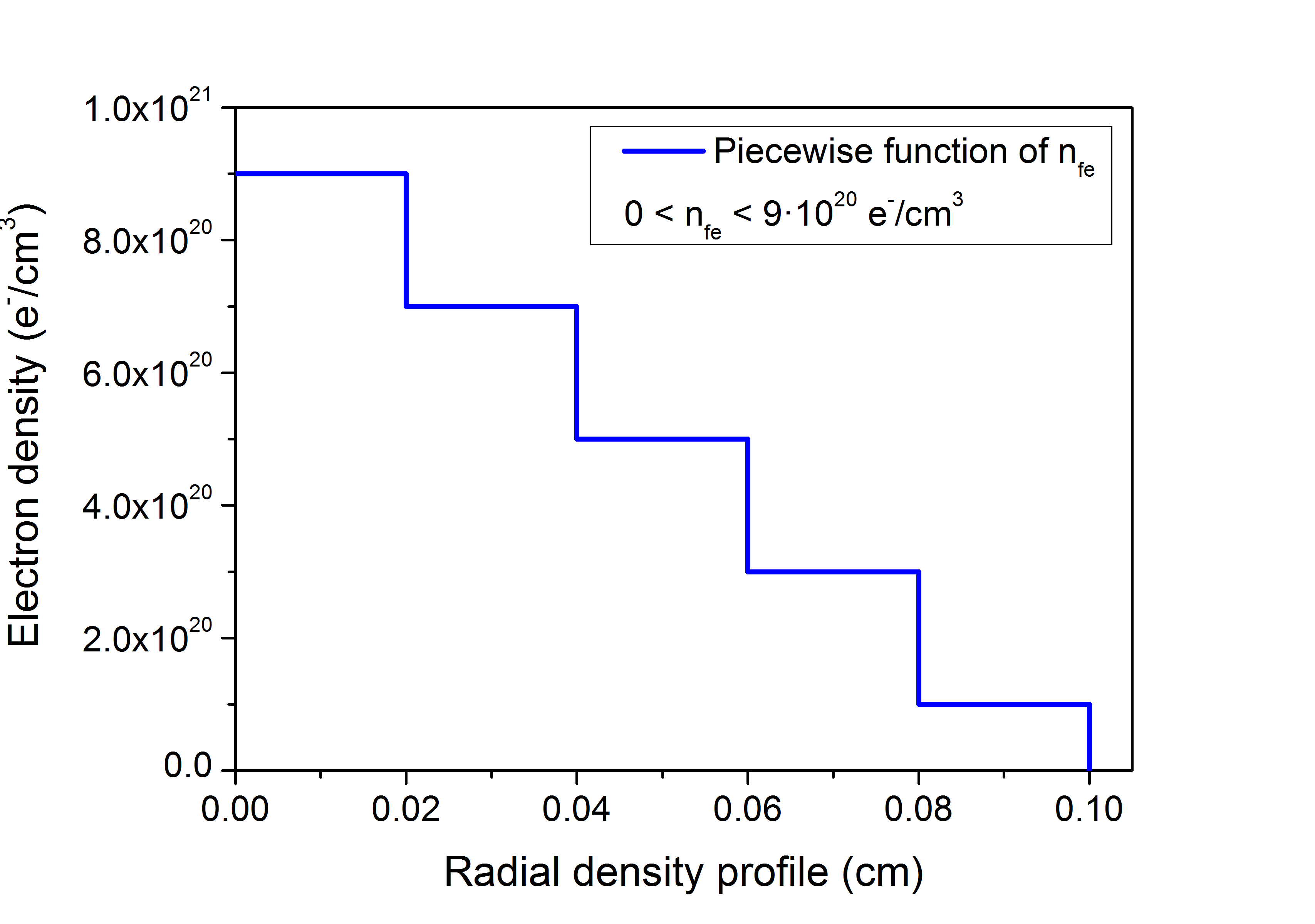} 
\caption{Electron density as a piecewise function of the radius.}
\label{F06}
\end{figure}

The energy loss of a proton beam has been also considered for a plasma with radial symmetry. In this case a rising electron density from external shells to inner core has been simulated by means of piecewise function, as it can see in Figure \ref{F06}.

\begin{figure}
\centering
\includegraphics[width=\linewidth]{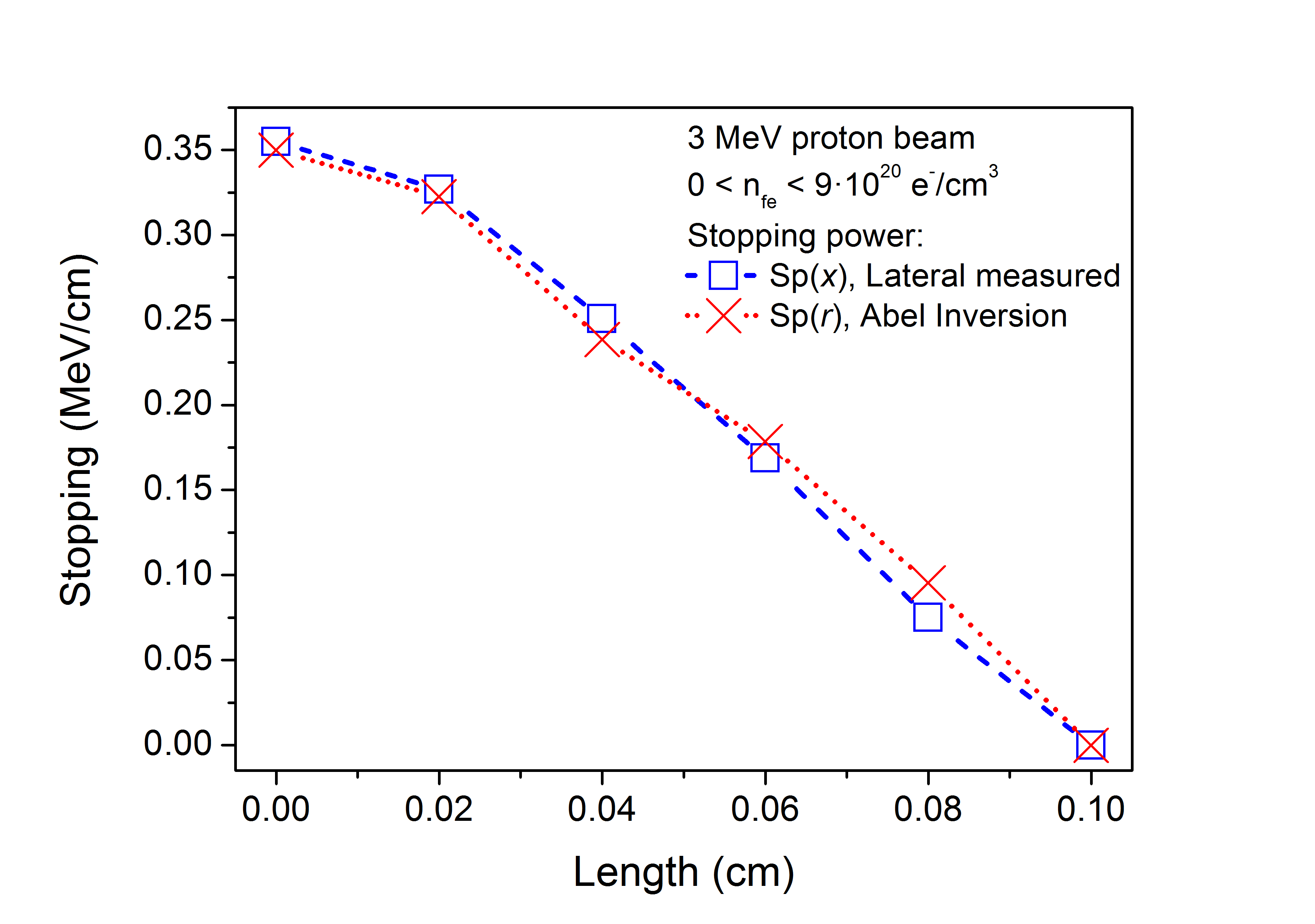} 
\caption{Stopping power as a function of lateral and radial points set.}
\label{F07}
\end{figure}

The energy loss of a proton beam is obtained in every $x_{i}$. Then, $Sp(x_{i})$ can be calculated, and applying the Abel inversion to this lateral point set, by means of Nestor-Olsen method, it is possible to obtain the stopping as function of the radius, $Sp(r_{i})$, as it is showed in Figure \ref{F07}.

\section{Conclusions}\label{Conclusions}

The stopping power of partially ionized plasma has been divided into two contributions. For free electrons, the RPA dielectric function obtained from interpolated values of discrete points of RPA calculations has been proposed. It has been proved that the differences between the calculated RPA and the interpolated one are negligible. In the case of bound electrons, a set of formulas for high and low projectile velocities has been proposed, which have the advantage to result in positive values of stopping for any proton velocity.

The energy loss has been evaluated using an iterative scheme, that provides an accurate value of Bragg peak and total depth stopping for a proton beam that traverses an extended plasma. In the case of plasma with radial symmetry, the Abel inversion has been used to obtain radial parameters from lateral measurements.

Finally, two kinds of plasma has been analyzed using the previous methods. The first one, created by a laser prepulse, is approximated by a piecewise function and compared with a rectangular profile with the same particle number quantity. In both cases, the final energy loss is practically the same, but showing some differences in the proton beam energy profile. The second kind of plasma has a symmetrical radial distribution, with a density that decreases from inside to outside. The Abel inversion has been applied to the stopping estimated from lateral measurement, obtaining the stopping as function of radius which is more closer to the radial electron density proposed.

\begin{acknowledgements}
This work is supported by the Spanish Ministerio de Econom\'ia y Competitividad (under Project MINECO: ENE2013-45661-C2-1-P). I would like to acknowledge University of Castilla-La Mancha and Diputaci\'on of Ciudad Real by provide me the actual doctoral grant. I would also like to express my gratitude to C\'atedra ENRESA that supports me with a doctoral stay in the Max Born Institute in Berlin, and to my supervisors there for their invitation and kindly treatment.

\end{acknowledgements}

\bibliographystyle{actapoly}
\bibliography{biblio}

\end{document}